\documentclass[12pt,preprint]{aastex}

\newcommand {\Ci} {\rm C\,{\small{I}}\ }
\newcommand {\Fei} {\rm Fe\,{\small{I}}\ }
\newcommand {\Oi} {\rm O\,{\small{I}}\ }
\newcommand {\kms} {\,{\rm km\,s}^{-1}}

\shorttitle{Fast rotators in M4}
\shortauthors{Lovisi et al.}
 
\begin{document} 
\title{Fast rotating Blue Stragglers in the globular cluster
  M4\footnote{Based on observations collected at the ESO-VLT under
    program 081.D-0356.  Also based on observations collected at the
    ESO-MPI Telescope under program 69.D-0582, and with the NASA/ESA
    {\it HST}, obtained at the Space Telescope Science Institute,
    which is operated by AURA, Inc., under NASA contract NAS5-26555.}
}

\author{
L. Lovisi\altaffilmark{2},
A. Mucciarelli\altaffilmark{2}, 
F.R. Ferraro\altaffilmark{2},
S. Lucatello\altaffilmark{3,4},
B. Lanzoni\altaffilmark{2}, 
E. Dalessandro\altaffilmark{2}, 
G. Beccari\altaffilmark{5},
R.T. Rood\altaffilmark{6},
A. Sills\altaffilmark{7},
F. Fusi Pecci\altaffilmark{8},
R. Gratton\altaffilmark{3},
G. Piotto\altaffilmark{9}
}
\affil{\altaffilmark{2} Dipartimento di Astronomia, Universit\`a degli Studi
di Bologna, via Ranzani 1, I--40127 Bologna, Italy}
\affil{\altaffilmark{3} INAF- Osservatorio Astronomico di Padova,
  Vicolo dell'Osservatorio 5, I--35122 Padova, Italy}
\affil{\altaffilmark{4} Excellence Cluster Universe, Technische
  Universit\"at M\"unchen, Boltzmannstr. 2, D--85748, Garching,
  Germany}
\affil{\altaffilmark{5}
ESA, Space Science Department, Keplerlaan 1, 2200 AG Noordwijk, Netherlands} 
\affil{\altaffilmark{6}Astronomy Department, University of Virginia, 
P.O. Box 400325, Charlottesville, VA, 22904,USA}
\affil{\altaffilmark{7}Department of Physics and Astronomy, McMaster University, 
 1280 Main Street West, Hamilton, ON, L8S 4M1, Canada}
\affil{\altaffilmark{8}
INAF- Osservatorio Astronomico di Bologna, Via Ranzani, 1, 40127 Bologna, Italy}
\affil{\altaffilmark{9}
 Dipartimento di Astronomia, Universit\`a di Padova, 
 Vicolo dell'Osservatorio 5, I--35122 Padova, Italy} 
\date{1 july, 10}
 
\begin{abstract} 
We have used high resolution spectra obtained with the spectrograph
FLAMES at the ESO Very Large Telescope to determine the kinematical
properties and the abundance patterns of 20 blue straggler stars
(BSSs) in the globular cluster M4. We found that $\sim 40\%$ of the
measured BSSs are fast rotators (with rotational velocities $> 50
\kms$). This is the largest frequency of rapidly rotating BSSs ever
detected in a globular cluster.  In addition, at odds with what has
been found in 47 Tucanae, no evidence of carbon and/or oxygen
depletion has been revealed in the sample of 11 BSSs for which we were
able to measure the abundances.  This could be due either to low
statistics, or to a different BSS formation process acting in M4.
\end{abstract} 
 
\keywords{blue stragglers; globular clusters: individual (NGC 6121);
  stars: abundances; stars: evolution; techniques: spectroscopic}

\section{Introduction} 
Blue straggler stars (BSSs) are brighter and bluer (hotter) than the
main sequence (MS) turnoff and they mimic a rejuvenated stellar
population in globular clusters (GCs).  From their position in the
colour-magnitude diagram (CMD) and from direct measurements (Shara et
al. 1997; see also Ferraro, Fusi Pecci \& Bellazzini 1995), they are
known to be more massive than the normal MS stars, thus indicating
that some process able to increase the initial mass of single stars
must be at work: BSSs could be generated by collision-induced stellar
mergers (COL-BSSs; Hills \& Day 1976), or they may form by the
mass-transfer activity in a binary system (MT-BSSs; McCrea 1964),
possibly up to the complete coalescence of the two companions. BSSs in
different environments could have different origins (Fusi Pecci et
al. 1992). In particular, BSSs in loose GCs might be produced from the
coalescence of primordial binaries, whereas in high-density GCs BSSs
might arise mostly from stellar collisions, particularly those
involving binaries. Moreover, there is evidence that both mechanisms
could act simultaneously within the same cluster (see the two distinct
sequences of BSSs recently discovered in M30 by Ferraro et al. 2009),
also depending on the different stellar densities at various distances
from the cluster center. This is suggested by the bimodality of the
BSS radial distribution detected in several GCs (Ferraro \& Lanzoni
2009 for a review).  While theoretical predictions about the
properties of BSSs generated by different formation channels are still
uncertain and controversial, the search for chemical patterns onto the
BSS surface seems to be the most promising route to discriminate
between the two scenarios. In fact, hydrodynamic simulations (Lombardi
et al. 1995; Sarna \& DeGreve 1996) suggest that very little mixing
should occur between the inner cores and the outer envelopes of the
colliding stars (hence COL-BSSs should not show any abundance
anomaly), while depleted surface abundances of carbon (C) and oxygen
(O) are expected for MT-BSSs, since the accreted material should come
from the core region of a peeled parent star where nuclear processing
has occurred.  Indeed this seems to be the case for the sub-sample of
BSSs with significant CO-depletion discovered in 47 Tucanae (47Tuc) by
Ferraro et al. (2006, hereafter F06).
 
In this framework, we are conducting an extensive survey of surface
abundance patterns and kinematical properties of BSSs in a selected
sample of GCs, by using the Very Large Telescope (VLT) of the European
Southern Observatory (ESO).  In this letter we report on new findings
of this project concerning the GC M4.

\section{Observations}
The observations were performed at the ESO-VLT during three nights in
June 2008, using the multi-object facility FLAMES-GIRAFFE (Pasquini et
al. 2002). The sample includes 20 BSSs, 53 subgiant branch stars
(SGBs) and 38 red giant branch stars (RGBs). The spectroscopic target
selection has been performed on a photometric catalog obtained by
combining ACS@HST data for the central region (i.e., at radial
distances $r<100\arcsec$) and WFI@ESO observations for the outer
region. We have also taken into account the stellar proper motions for
the wide-field sample (Anderson et al. 2006), and discarded all the
stars having a source with a comparable or a brighter luminosity at a
distance smaller than $3\arcsec$. The selected BSS sample represents
$\sim 70\%$ of the entire population within $800\arcsec$ ($\approx 11$
core radii) from the cluster centre (Lanzoni et al. 2010, in
preparation), with $\sim 50\%$ of them being located at
$r<100\arcsec$.  The RGBs and SGBs have been selected from the WFI
sample at $100\arcsec< r<800\arcsec$.
 
Three different setups were used for the spectroscopic observations:
HR15, HR18 and HR22, suitable to sample the H$\alpha$ line, the \Oi
triplet at $\lambda \simeq 7774$ \AA, and the \Ci line at $\lambda
\simeq 9111$ \AA, respectively. Exposure times amount to one hour for
the HR15 setup, and two hours each for the HR18 and HR22.  Spectra
have been pre-reduced using the standard GIRAFFE ESO pipeline. The
accuracy of the wavelength calibration has been checked by measuring
the position of a number of emission telluric lines (Osterbrock et
al. 1996).  Then, we subtracted the mean sky spectrum from each
stellar spectrum. By combining the exposures, we finally obtained
median spectra with signal-to-noise ratios S/N$\simeq 50-100$ for the
selected BSSs and SGBs, and S/N$\simeq 100-300$ for the RGBs.

\section{Analysis and results}
The procedure adopted to derive the radial and rotational velocities,
and the [O/Fe], [C/Fe] and [Fe/H] abundance ratios for the observed
sample is summarized below.  Table \ref{bss} lists the values obtained
for the BSSs, together with the adopted temperatures and gravities.

\noindent
{\bf Radial velocities --} Radial velocities ($V_{\rm rad}$) were
measured using the IRAF task \textit{fxcor} that performs the Fourier
cross-correlation between the target spectra and a template of known
radial shift, following the prescriptions by Tonry \& Davis (1979). As
a template for the samples of BSSs, SGBs and RGBs we used the
corresponding spectra with the highest S/N ratio, for which we
computed $V_{\rm rad}$ by measuring the wavelength position of a few
tens of metallic lines. Radial velocity values obtained from the three
different setups are consistent each other within the uncertainties
which are of the order of $0.5\kms$ for the SGB stars and most of the
BSSs (see Table 1), and $0.15\kms$ for the RGBs.  For the fast
rotating stars (see below) $V_{\rm rad}$ has been estimated from the
centroid of the H$\alpha$ line, which is almost unaffected by
rotation.

\begin{figure}
\plotone{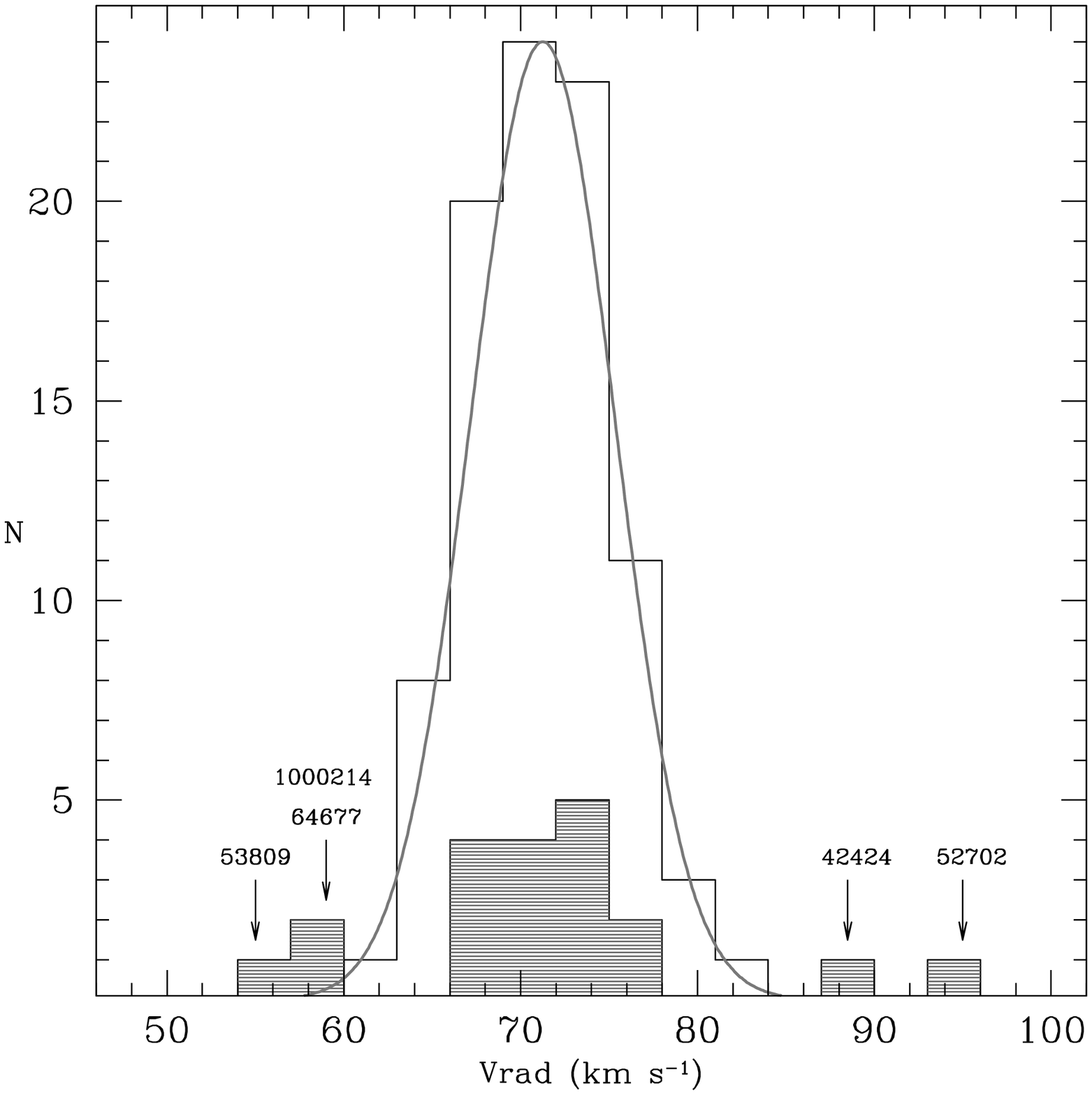}
\caption{Radial velocity distribution for the observed stars in
  M4. The empty histogram refers to SGBs and RGBs, the shaded
  histogram to BSSs. The solid curve is the gaussian best-fit to the
  giant $V_{\rm rad}$ distribution. BSSs displaying discordant radial
  velocities with respect to the cluster distribution are labelled.}
\label{vradisto}
\end{figure}

Figure \ref{vradisto} shows the derived $V_{\rm rad}$ distribution for
the giants (RGBs+SGBs) and the BSSs. The mean radial velocity of the
total sample is $\langle V_{\rm rad}\rangle = 71.28\pm0.50\kms$, with
a dispersion $\sigma=5.26 \kms$. The distribution for the giant stars
is peaked at nearly the same value, $\langle V_{\rm rad}\rangle =
71.25\pm0.43\ (\sigma=4.08)\ \kms$, which we adopt as the systemic
velocity of M4. This is in good agreement with previous determinations
(Peterson et al. 1995, Harris 1996, Marino et al. 2008, Sommariva et
al. 2009, Lane et al. 2009).  The average of the BSS radial velocity
distribution is also similar, $\langle V_{\rm rad}\rangle
=71.40\pm2.01\kms$, but the dispersion is larger ($\sigma=9\kms$) due
to the discordant values measured for five of these stars (see Table
1).

\noindent
{\bf Rotational velocities --} To derive the stellar rotational
velocities we have used the method described by Lucatello \& Gratton
(2003). In particular we have computed the Doppler broadening due to
the rotation of the object ($I_{\rm rot}$), which is linked to the
rotational velocity ($v_{\rm rot} \sin i$) by means of a coefficient
$\alpha$: $I_{\rm rot} = v_{\rm rot} \sin i/\alpha$ (see equation 3 in
Lucatello \& Gratton 2003, where $I_{\rm rot}$ is indicated as
``$r$''). In principle, the coefficient $\alpha$ should be calibrated
by using rotational velocity standards observed with the same
instrumental configuration.  Although we have not performed such a
calibration, the comparison with synthetic spectra computed for
different rotational velocities indicates that $\alpha$ is close to
unity. We have adopted $1.5\pm 0.5\kms$ for the micro-turbulent
velocity\footnote{While this value has been strictly measured for a
  few SGB stars only, the impact of such an assumption on the derived
  rotational velocities and chemical abundances is negligible.}, and
the relation obtained by Gray (1992) for the
macro-turbulence. Temperatures have been derived by fitting the wings
of the H$\alpha$ line, which are insensitive to other parameters like
metallicity and gravity (Fuhrmann 1993). In particular, we performed a
$\chi^2$ minimization between the observed H$\alpha$ profile and a
grid of synthetic spectra computed with different temperatures, by
using SYNTHE code (Kurucz 1993).  Typical uncertainties for the
temperatures are of the order of $\sim 50$ K for the BSSs, and $\sim
100-150$ K for the giants.  Since the adopted technique is efficient
only for rotational velocities up to $\simeq 50 - 60\kms$, the values
of $I_{\rm rot}$ for faster stars have been estimated by comparing the
observations in the \Oi triplet region with rotationally broadened
synthetic spectra.

\begin{figure}
\plotone{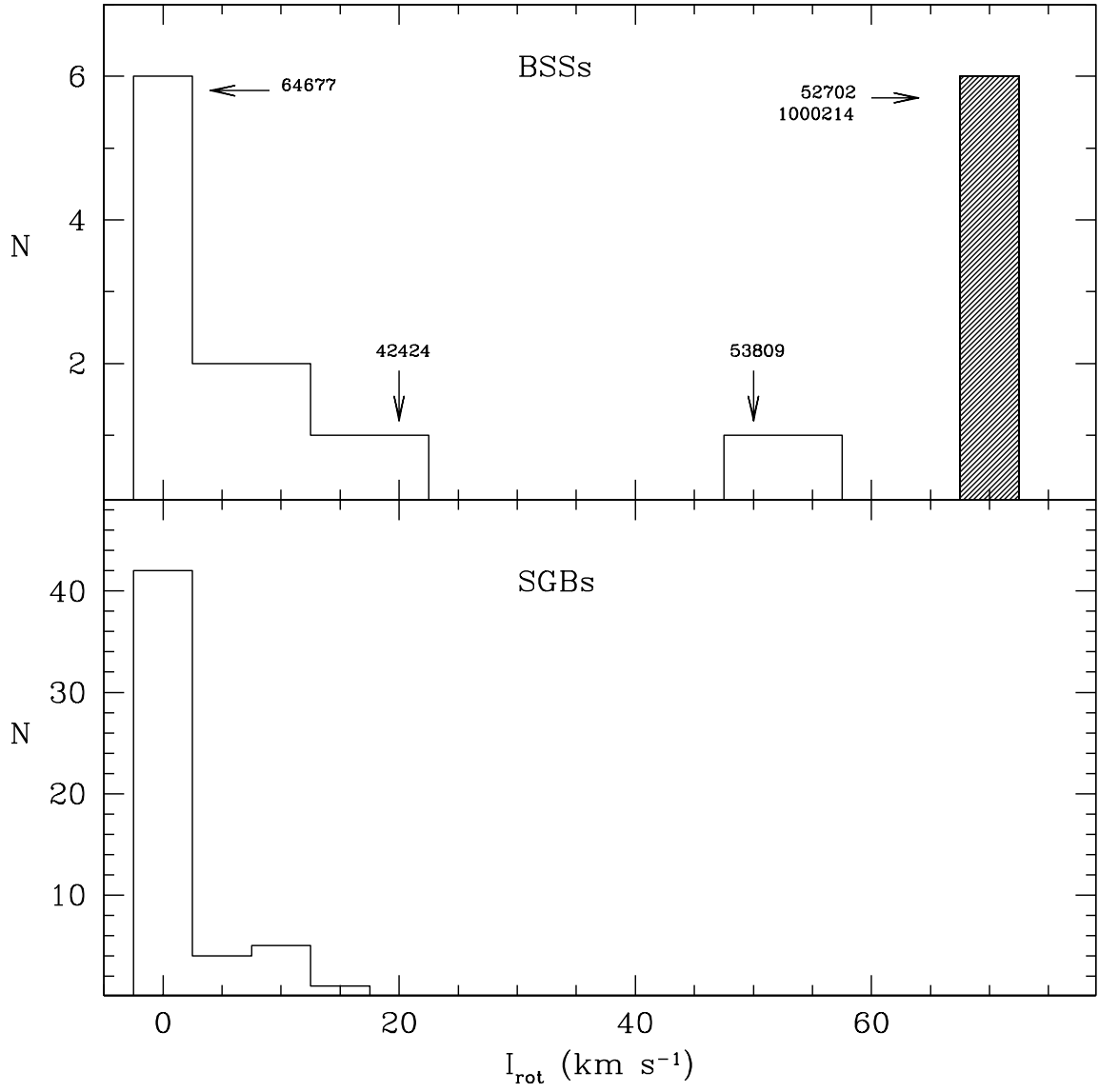} 
\caption{Rotational index distribution derived for the observed BSSs
  (upper panel) and SGB stars (lower panel). The six BSSs with $I_{\rm
    rot}\ge 70 \kms$ are all plotted within shaded block. The five
  BSSs with anomalous $V_{\rm rad}$ are labelled. All RGBs have
  $I_{\rm rot}=0\kms$ (Mucciarelli et al. 2010, ApJ submitted) and are
  not shown for clarity.}
\label{rotisto}
\end{figure}

Figure \ref{rotisto} presents the derived rotational index
distributions: that of the giants is peaked at $I_{\rm rot}=0.0 \kms$,
with the highest value being $13.4\pm 3.4\kms$ for a SGB.  The
rotational index distribution for the BSSs (see Table 1) is quite
different, with eight stars (40\% of the total) being fast rotators,
i.e., rotating at more than $50\kms$ (while normal F-G type stars
typically spin at less than $\sim 30\kms$; Cort\'es et
al. 2009). Interestingly, three (out of five) BSSs with anomalous
$V_{\rm rad}$ also are fast rotators.

\noindent
{\bf Chemical abundances --} As the reference population needed to
identify possible anomalies in the BSS surface abundances, we have
considered the SGBs, since episodes of mixing and dredge-up may have
modified the primordial abundance patterns in the RGBs. Chemical
abundances have been derived from the equivalent width measurements by
using the WIDTH9 code (Kurucz 1993; Sbordone et al. 2004). Gravities
have been determined (within 0.2 dex) by comparing the target position
in the CMD with a grid of evolutionary tracks extracted from the
\textit{BaSTI Library} (Pietrinferni et al. 2006)\footnote{The
    derived values for BSSs may be affected by systematic
    uncertainties, since these stars could be underluminous for their
    masses (van der Bergh et al. 2001; Sandquist et al. 2003; Mathieu
    \& Geller 2009).  While we do not have direct spectroscopic
    indicators for gravity, uncertainties of 0.2 dex translate in
    abundances variations smaller than 0.1 dex.}.  This also yielded
to a mass distribution for the observed BSSs, which peaks at $\sim 1
M_\odot$, with the most massive object being at $\sim 1.3 M_\odot$.
Abundance errors have been computed by taking into account the
uncertainties on the atmospheric parameters and those on the
equivalent width measurements. For each star they typically are of the
order of $0.1-0.2$ dex.

The iron content for the SGBs and BSSs has been derived from the
equivalent widths of about ten and 2--7 \Fei lines, respectively.  For
the SGBs the resulting average iron abundance is [Fe/H]$= -1.10 \pm
0.01$, with a dispersion $\sigma = 0.07$ about the mean, in good
agreement with previous values (ranging between $-1.20$ and $-1.07$;
Harris 1996; Ivans et al. 1999; Marino et al. 2008; Carretta et
al. 2009).  Because of the significant deformation of the spectral
line profiles, no iron abundance has been derived for the eight fast
rotators. Moreover, technical failures in the spectrograph fiber
positioning prevented us to measure it for two additional objects (see
Table 1).  The iron abundance obtained for the remaining ten BSSs has
a mean value of $-1.27$ and a dispersion $\sigma=0.10$, consistent,
within the errors, with the values derived for the SGBs.

\begin{figure}
\plotone{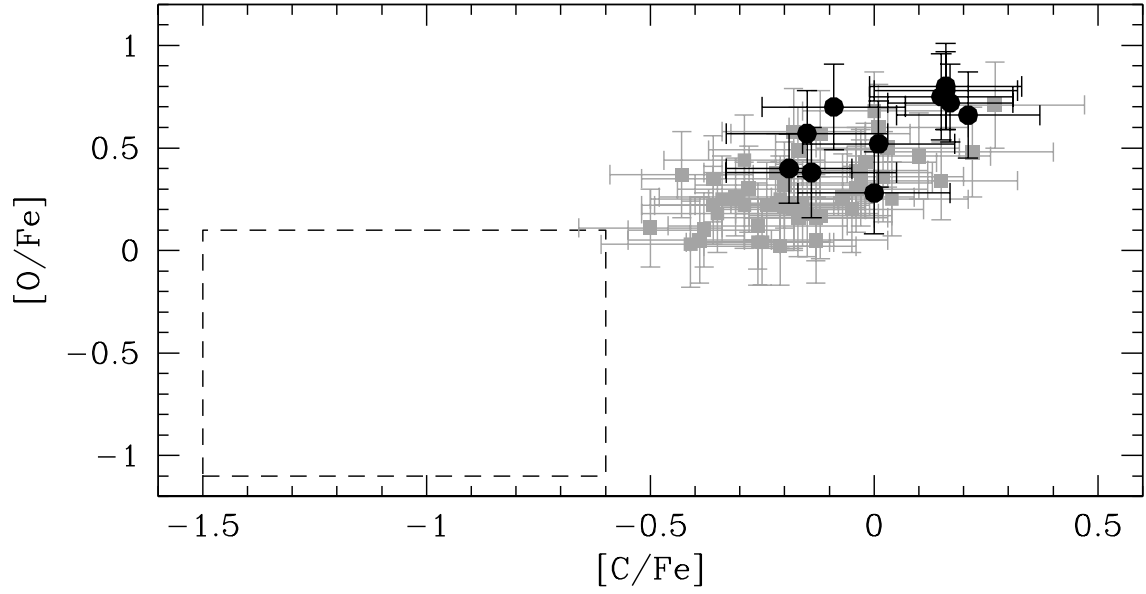} 
\caption{[O/Fe] ratio as a function of [C/Fe] for the measured BSSs
  (black dots) and SGBs (gray dots). The dashed box marks the position
  of the CO-depleted BSSs observed in 47Tuc (Ferraro et al. 2006).}
\label{abb}
\end{figure}

Oxygen and carbon abundances have been computed from the equivalent
widths of the \Oi triplet and the \Ci lines respectively, and the
derived values have then been corrected for non-local thermodynamic
equilibrium effects. For O abundances these corrections were derived
by interpolating the grid by Gratton et al. (1999); for C abundances
we adopted the empirical relation obtained by interpolating the values
listed by Tomkin et al. (1992).  The resulting average abundances for
the SGB sample are [C/Fe]$=-0.16\pm 0.02$ ($\sigma = 0.17$) and
[O/Fe]$=0.29\pm 0.02$ ($\sigma = 0.17$). No measurements have been
possible for the very fast rotating BSSs and for a few other objects
(see Table 1). Hence, we were able to measure both the C and O
abundances only for 11 BSSs, out of 20 observed.  Figure \ref{abb}
shows the results obtained in the [C/Fe]$-$[O/Fe] plane.  The values
measured for the 11 BSSs are in agreement with those of the SGBs, with
no evidence of depletion either in carbon or in oxygen. We finally
notice that also in the 3 cases for which only the oxygen \textit{or}
the carbon abundance has been measured (see Table 1), the values
obtained are in agreement with those of the SGBs.

\section{Discussion}
Before discussing in details the main findings of the present work it
is necessary to verify whether some of the investigated stars do not
belong to the cluster.  In particular, five BSSs have been found to
display anomalous $V_{\rm rad}$, which may cast some doubts about
their membership.  However, BSSs \#42424 and \#64677 have measured
proper motions well in agreement with those of the cluster members
(Anderson et al. 2006). All the other objects have measured rotational
velocities significantly larger than expected for normal stars of the
same spectral type, thus making unlikely that they belong to the
field.  We have also used the Besan\c{c}on Galactic model (Robin et
al. 2003) to derive the radial velocity and metallicity distributions
of the Galactic field stars in the direction of M4, within the same
magnitude and colour intervals shared by our BSS sample.  The $V_{\rm
  rad}$ distribution is peaked at $-14.6\kms$ and has a dispersion
$\sigma = 50.7\kms$. As a consequence, the probability that the BSSs
with anomalous radial velocity belong to the field is always smaller
than 1.7\%.  The theoretical metallicity distribution of field stars
is peaked at [Fe/H]$=-0.17\pm 0.02$ ($\sigma = 0.45$). The iron
abundance measured in two of the five BSSs with anomalous radial
velocity ([Fe/H]$=-1.35$ for \#42424, and $-1.23$ for \#64677) clearly
is largely inconsistent with the field value and concordant, within
the errors, with that of M4 stars.  Based on these considerations we
therefore conclude that all the BSSs with anomalous $V_{\rm rad}$ are
indeed members of M4.  We notice that if the discrepancies were caused
by the orbital motion in binary systems, under realistic assumptions
about the total mass, the orbits would be reasonable for a GC, with
separations ranging from a few to 10--20 AU.  These BSSs do not show
any evidence of photometric variability (Kaluzny et al. 1997), and
variations of $V_{\rm rad}$ with full amplitudes exceeding $3\kms$ on
a time interval of 72 hours are excluded by our observations for BSSs
\#42424 and \#64677 (no such information is available for the fast
rotators). However, these results do not disprove that they
  are in binary systems. For instance, they are still consistent with
  what expected for $\sim 90$\% of binaries characterized by an
  eccentricity-period distribution similar to that recently observed
  by Mathieu \& Geller (2009) and populating the tail of the velocity
  distribution in M4 (Mathieu \& Geller, private
  communication). Indeed, further observations are urged to search for
  clear-cut signatures of binarity.

The fact that none of the 14 BSSs for which we measured C and/or O
abundances shows signatures of depletion is quite intriguing.  Out of
the 42 BSSs investigated in 47Tuc, F06 found that 6 (14\%) are
C-depleted, with 3 of them also displaying O-depletion. Accordingly,
in M4 we could have expected 1-2 BSSs with depleted carbon abundance,
and 0-1 BSS with both C and O depletion.  Hence, the resulting
non-detection may just be an effect of low statistics and is still
consistent with the expectations.  Alternatively the lack of chemical
anomalies in M4 BSSs might point to a different formation process:
while at least 6 BSSs (the CO-depleted ones) in 47Tuc display surface
abundances consistent with the MT formation channel, all the
(investigated) BSSs in M4 may derive from stellar collisions, for
which no chemical anomalies are expected. Finally, it is also possible
that the CO-depletion is a transient phenomenon (F06) and (at least
part of) the BSSs in M4 are indeed MT-BSSs which already evolved back
to normal chemical abundances.

\begin{figure}
\plotone{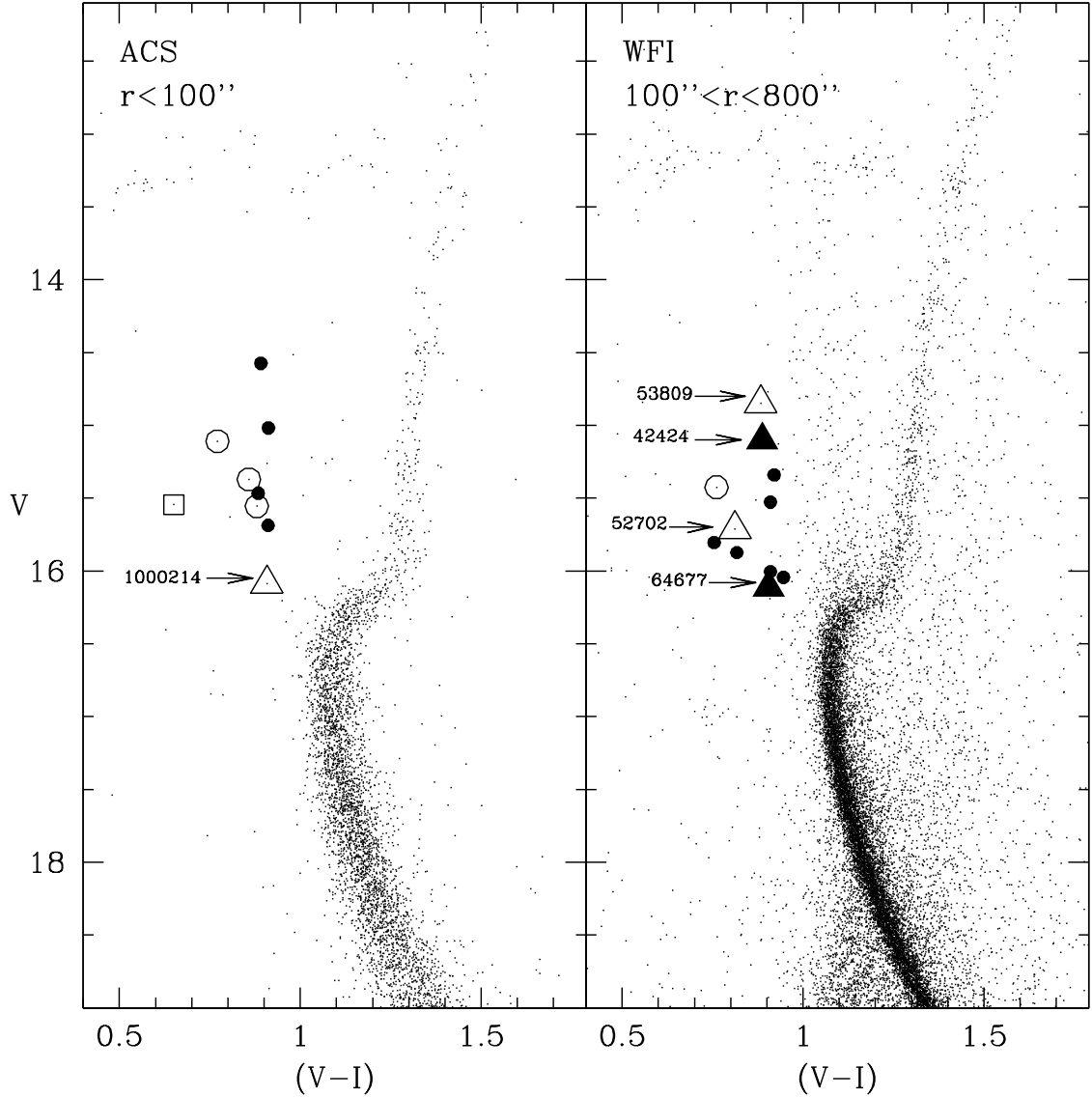} 
\caption{CMD of M4 for the ACS and the WFI datasets. The observed BSSs
  are highlighted: large black dots mark normal BSSs, triangles mark
  BSSs with anomalous $V_{\rm rad}$, the square indicates the W Uma
  star, open symbols mark fast rotators ($I_{\rm rot}> 50\kms$).}
\label{cmd}
\end{figure}

The most intriguing result of this study is the discovery that a large
fraction (40\%) of the investigated BSSs in M4 are fast rotators, with
rotational velocities ranging from $\sim 50\kms$, up to more than
$150\kms$.  We emphasize that this is the largest population of fast
rotating BSSs ever found in a cluster. Approximately $30\%$ of the
BSSs spinning faster (at $20-50\kms$) than MS stars of the same colour
has been recently found in the old open cluster NGC188 (Mathieu \&
Geller 2009), while BSSs in younger open clusters are found to rotate
slower than expected for their spectral type (e.g., Shetrone \&
Sandquist 2000; Sch\"{o}nberner et al. 2001).  For GCs only scarce and
sparse data have been collected to date. The most studied case is that
of 47Tuc, where 3 (7\%) BSSs out of the 45 measured objects (Shara et
al. 1997; De Marco et al. 2005; F06) have rotational velocities larger
than $50\kms$, up to $\sim 155\kms$.  The object studied by Shara et
al. (1997) is the second brightest BSS in 47Tuc, and all the others
are located at the low-luminosity end of the BSS region in the CMD. In
addition they span almost the entire range of surface temperatures and
distances from the cluster centre. For comparison, apart from being
more numerous, the fast rotating BSSs in M4 are also found at all
luminosities, temperatures and radial distances (see Figure
\ref{cmd}).  There is a weak indication that the rapid rotating BSSs
in M4 tend to be more centrally segregated than normal BSSs, even if
the small number of stars in our sample prevents a statistically
robust result (following a Kolmogorov-Smirnov test, the probability
that the two distributions are extracted from the same parent
population is $\sim 44\%$).  The fastest rotating BSS in M4
(\#2000121, with $I_{\rm rot} \sim 150-200\kms$) corresponds to star
V53 of Kalunzy et al. (1997), which is classified as a W UMa contact
binary. Interestingly, even in the F06 sample of 47Tuc the fastest BSS
(spinning at $\sim 100\kms$) is a W UMa binary. These two stars also
occupies a very similar position in the CMD, at the high-temperature
and low-luminosity side of the BSS region.

Unfortunately, from the theoretical point of view, rotational
velocities cannot be easily interpreted in terms of BSS formation
processes. In fact, fast rotation is expected for MT-BSSs because of
angular momentum transfer, but some braking mechanisms may then
intervene with efficiencies and time-scales that are still unknown.
The predictions about rotational velocity of COL-BSS are controversial
(Benz \& Hills 1987; Leonard \& Livio 1995; Sills et al. 2005). In
particular, the latter models show that angular momentum losses
through disk locking are able to decrease the BSS rotational velocity
from values as high as $\sim 100\kms$ to $20\kms$. Hence, the fast
rotating BSSs observed in M4 could be generated either through mass
transfer activity or through stellar collisions, on condition that no
significant braking has (still) occurred.

Three of the BSSs (namely \#52702, \#53809 and \#1000214) of the five
with anomalous $V_{\rm rad}$ also are fast rotators. Such a high
rotation is difficult to account for by synchronization or mass
transfer effects in binary systems, since the orbital separation would
not be small enough.  A fascinating alternative is that such anomalies
in the radial and rotational velocities are due to three- and
four-body interactions that occurred in the cluster core: these could
have originated fast spinning BSSs and kicked them out to the external
regions at high speed (apart from star \#1000214, the other two are
located well beyond the cluster core radius).  Interestingly enough,
the dynamical history of the cluster could support such a scenario. In
fact, although its stellar density profile is well reproduced by a
King model, recent Monte-Carlo simulations (Heggie \& Giersz 2008)
suggest that M4 could be a post-core collapse cluster, its core being
sustained by the ``binary burning'' activity. The fast rotating and
high velocity BSSs could be the signature of such an activity.

While bringing a wealth of information, the results obtained to date
for 47Tuc and M4 are still too scarce to provide a clear overall
picture.  Collecting additional data on rotational velocities and
chemical abundances for a significant number of BSSs in a sample of
GCs with different structural parameters is indeed a crucial
requirement for finally understanding the formation mechanisms of
these puzzling stars and their link with the cluster dynamical
history.
 
\acknowledgements We acknowledge R. Bedin and Y. Momany for useful
discussions, and the referee, Robert D. Mathieu, for helpful
suggestions in improving the paper.  This research was supported by
the Agenzia Spaziale Italiana (under contract ASI-INAF I/016/07/0), by
the Istituto Nazionale di Astrofisica (INAF, under contract
PRIN-INAF2008) and by the Ministero dell'Istruzione, dell'Universit\`a
e della Ricerca.

\begin{deluxetable}{rccrrrrrl} 
\tablecolumns{9} 
\tablewidth{0pc} 

\tablecaption{Identification numbers, temperatures, gravities, radial
  velocities, rotational indices, [O/Fe], [C/Fe] and [Fe/H] abundance
  ratios for the observed BSSs. ``RVa'' and ``FR'' flag the stars with
  anomalous radial velocities and high rotational speed ($I_{\rm
    rot}>50\kms$), respectively.}
\tablehead{
  id & T & $\log(g)$ & $V_{\rm rad}~~~~$ & $I_{\rm rot}~~~~$& [O/Fe] & [C/Fe] & [Fe/H] & Notes\\  
     &(K)&           &  ($\kms$)    & ($\kms$)    &         &        &        &       
}
\startdata 
  31209	&  7300 & 4.3 & $74.6\pm 0.6$ & $  0.0\pm 2.0$ & $0.70$ & $-0.09$ & $ -1.18$ &         \\ 
  37960	&  6650 & 4.0 & $72.4\pm 0.6$ & $  2.0\pm 1.3$ & $0.52$ & $ 0.01$ & $ -1.19$ &         \\
  42424	&  6850 & 3.8 & $87.8\pm 0.9$ & $ 22.3\pm 0.2$ & $0.80$ & $ 0.16$ & $ -1.35$ & RVa     \\  
  43765	&  6950 & 4.1 & $66.3\pm 6.9$ & $ 56.8\pm 0.1$ & $0.78$ & $ 0.16$ & $   -  $ & FR      \\
  44123	&  6900 & 4.2 & $71.2\pm 4.6$ & $ 10.3\pm 0.4$ & $  - $ & $ 0.10$ & $   -  $ &         \\
  47856	&  6450 & 4.1 & $66.1\pm 0.6$ & $  0.0\pm 2.5$ & $0.57$ & $-0.15$ & $ -1.20$ &         \\
  47875	&  6550 & 3.9 & $71.3\pm 0.5$ & $  0.0\pm 7.8$ & $0.73$ & $   - $ & $ -1.37$ &         \\
  52702	&  6900 & 4.2 & $95.1\pm 7.8$ &   70 - 120     & $  - $ & $   - $ & $   - $  & FR; RVa \\
  53809	&  6700 & 3.8 & $55.9\pm 6.9$ & $ 51.6\pm 0.1$ & $0.75$ & $ 0.15$ & $   -  $ & FR; RVa \\
  54714	&  6500 & 4.1 & $74.5\pm 0.6$ & $  0.0\pm 6.4$ & $0.28$ & $ 0.00$ & $ -1.20$ &         \\
  64677	&  6500 & 4.2 & $58.6\pm 0.6$ & $  0.0\pm 2.0$ & $0.38$ & $-0.14$ & $ -1.23$ & RVa     \\
1000070	&  6550 & 3.7 & $67.4\pm 0.4$ & $  4.3\pm 0.7$ & $0.72$ & $ 0.17$ & $ -1.23$ &         \\
1000117	&  6700 & 4.0 & $72.2\pm 4.6$ &     70 - 100   & $  - $ & $   - $ & $   -  $ & FR      \\
1000125	&  7100 & 4.0 & $69.5\pm 6.9$ &     80 - 130   & $  - $ & $   - $ & $   -  $ & FR      \\ 
1000143	&  6650 & 4.0 & $68.4\pm 0.3$ & $  3.2\pm 0.9$ & $0.40$ & $-0.19$ & $ -1.29$ &         \\
1000214	&  6800 & 4.2 & $59.5\pm 5.9$ &     70 - 100   & $  - $ & $   - $ & $   -  $ & FR; RVa \\ 
2000075	&  6700 & 3.8 & $74.5\pm 0.5$ & $  8.3\pm 0.4$ & $0.66$ & $ 0.21$ & $ -1.45$ &         \\
2000085	&  7100 & 4.0 & $70.4\pm 7.8$ &     80 - 120   & $  - $ & $   - $ & $   -  $ & FR      \\ 
2000106	&  6750 & 4.0 & $76.1\pm 2.3$ & $ 12.6\pm 0.3$ & $  - $ & $ 0.10$ & $   -  $ &         \\
2000121	&  7350 & 4.3 & $76.4\pm 8.2$ &    150 - 200   & $  - $ & $   - $ & $   -  $ & FR; WUMa\\	
\hline
\enddata
\label{bss}
\end{deluxetable} 

\end{document}